\documentclass[floatfix,twocolumn,preprintnumbers,amsmath,amssymb,prb,superscriptaddress]{revtex4}
\usepackage{amsmath}
\usepackage{graphicx}%
\usepackage{dcolumn}
\usepackage{amsmath}
\usepackage{subfigure,hyperref}
\usepackage[normalem,normalbf]{ulem}

\usepackage{enumerate}

\newcommand{\be}{\begin{equation}}
\newcommand{\ee}{\end{equation}}

\usepackage{enumerate}

\begin{document}

\title{Hierarchical causality in financial economics}
\author{Diane Wilcox}
\email{diane.wilcox@wits.ac.za}
\affiliation{School of Computational and Applied Mathematics, University of the Witwatersrand, Johannesburg, South Africa}
\author{Tim Gebbie}
\email{tim.gebbie@physics.org}
\affiliation{School of Computational and Applied Mathematics, University of the Witwatersrand, Johannesburg, South Africa}
\begin{abstract}
 Hierarchical analysis is considered and a multilevel model is presented in order to explore causality, chance and complexity in financial economics. A coupled system of models is used to describe multilevel interactions, consistent with market data:    the lowest level is occupied by agents generating the prices of individual traded assets; the next level entails aggregation of stocks into markets; the third level combines shared risk factors with information variables and bottom-up, agent-generated structure, consistent with conditions for  no-arbitrage pricing theory;  the fourth level describes market factors  which originate in the greater economy and the highest levels are described by regulated market structure and the customs and ethics which define the  nature of acceptable transactions. A mechanism for emergence or innovation is considered and causal sources are discussed in terms of
 five causation classes.
\end{abstract}

\maketitle


\section{Introduction}

Coupled systems typically evolve under multiple sources of causality and chance. Market failures at the outset of the 21st century have highlighted that orthodox economics has not been able to provide models for stable pricing of investment securities. Multiple causes have been discussed for financial crises\cite{AG2007, OS2010, CFHGJKLS2009, CTSBCHK2012, HK2013} and it has become increasingly accepted that there are forms of causality other than those encompassed by the traditional reductionist views arising in physics, chemistry or economics \cite{S2000, vdBG2003, Minsky2008, Helbing2009}.  We address the causality question in financial economics by building a more causally complete view of real world systems. We commence with reviewing some of  key concepts.

Many authors have contemplated models of causality and Plato is attributed with claiming \cite{P1971}:

\vspace{-0.1cm}
\begin{quote}
Everything that becomes or changes must do so owing to some cause; for nothing can come to be without cause. \end{quote}

\vspace{-0.1cm}

Aristotle is documented as one of the very earliest theorists to delve into the {\em nature} of causation by describing four different kinds of causation, namely: material, formal, efficient and final \cite{Hulswit2002, GFRE2008}. Since then, ideas on causation have evolved. More recently, seminal directions include the works of  Simon \cite{S1952} et al, who
devised and improved methods for  determining sources of top-down causation in systems of differential equations and Campbell \cite{C1974} et al, whose focus stemmed from the analysis of  biological systems to highlight {\em selection} as a determinant for and against system features.
Advances in dynamical systems views have led to multi-scale modelling, where model development consistently takes into account empirical measurements at more than one scale \cite{VD1975,ZMA2005} and multi-layer networks analysts provide a formalism whereby distinguishable sub-networks are connected via multiple types of connections \cite{Sachs2012,DSCKMPGA2013,S2013}.

Stress-testing in the financial services typically entails the investigation of top-down and bottom-up causalities \cite{Sorge2004,AS2008,Haldane2009,BDT2014} acting on institutions. Hierarchical effects have been discussed with respect to the relationship between topological configuration of homogenous levels and information flow \cite{Helbing2009} and a graph-theoretic method for identifying clusters and hierarchical structure  has been developed by Di Matteo et al\cite{DiM2012}. Connectionist approaches to understanding causality relationships in markets have been addressed in the literature by Aubin \cite{Aubin2003}, where top-down causation is described by operators on product spaces, and Farmer \cite{F1990, F2002}, where  complex markets are deconstructed via an ecological systems perspective.

We investigate the applicability of a top-down causation scheme which is ordered by {\it coarse-graining} via the perspective  suggested by Auletta, Ellis and Jaeger \cite{AEJ2008, GFRE2008}. In this approach,
foundational features in heterogenous systems are identified and represented via averaged features, with averaging-scales  usually close to  prevailing scales of measurement for distinguishable  levels in the system.  Here, a hierarchy is represented by a system of coupled models which are applicable on different scales of interaction. It is the resulting loss of information, as one moves upward through scales,  that can lead to the identification of laws that function relatively autonomously at different levels. The independence of higher levels from the details of lower levels allows phenomenological laws to offer reasonable approximations at the higher levels for most observations of the system.

We are led to address the concept of causally complete models. Causally complete and causally closed models are considered in physics, where the latter refers to a perspective that the only kinds of causation in physical systems are physical and the former refers to the view that non-physical causes may be possible \cite{J2008}. In mathematics, completeness carries connotations of all things in the systems being attainable: (i) all theorems can be proved, (ii) limits of convergent sequences exist within the system  (iii) all objects in the system can be obtained as some representation of some core set, or (iv) all nodes in a network can be reached from every other nodes by a directed link.

G\"{o}del's Incompleteness Theorem exposes the reality that there will always exist unanswerable questions beyond the reach of any comprehensive  mathematical modelling framework. Analogously, there are typically  anomalies and  phenomena which will arise outside of any  attempt to model and regulate an economic system.  Thus, we consider the notion of causal completeness in the context of being able to {\em compare} complex models by  their ability to explain phenomena or quantities which can be observed or measured.

It is suggested by Ellis \cite{GFRE2008} that there are at least five different classes of top-down causation \cite{E2005,GFRE2008,CB2007}, listed here with slight variation in nomenclature:

\begin{description}
\item [TDC1]  Algorithmic top-down causation
\item [TDC2]  Top-down causation via non-adaptive information control
\item [TDC3]  Top-down causation via adaptive selection
\item [TDC4]  Top-down causation via feedback control of adaptive goals
\item [TDC5]  Top-down causation with adaptive selection of adaptive goals
\end{description}

In this paper, we present a simple hierarchical causality model which incorporates a plausible approximation of empirically measurable interactions, with a variety of top-down and bottom-up causality features, as a path to revising prevailing theoretical foundations in financial economics. To substantiate the model proposed, we provide a candidate for each of the distinct causation classes listed above, to give insight into the complex nature of causation and chance in financial markets. Parts of our hierarchical model remain consistent with existing arbitrage-free approaches to factor modelling \cite{WG2013b}.

To describe our model, we commence with a review of a seminal construct in the modelling of equity market causality. In arbitrage pricing theory, any (theoretically) riskless portfolio with an expectation of positive future returns must cost something now \cite{R1976,R2005}. Expected discounted returns can be estimated via  a linear factor model, based on market-wide risk premia and  historic,  (mostly) de-correlated stock-specific returns. In the factor model of Ross\cite{R1976,R2005}, risk premia which impact prices may be due to exogenous or endogenous contributions. Price changes which are different from noise are understood to be caused by top-down exogenous factors. Nevertheless, the resulting co-movements in prices may be describable  by endogenous factors within a given level in the hierarchy of complexity, due to efficient assimilation of exogenous information. In this model, systematic deviations from predictions provided by the explanatory factors cannot be sustained. Hence, over longer horizons the only return that a portfolio should yield will be a risk-free rate plus the risk premia associated with a finite set of common risk factors.

 It is generally accepted that real market dynamics are much less trivial. Deviations from prices predicated by prevailing models can arise. Such events may trigger trade based on an  expectation of profit due to detection of  a mispricing,  even though the anomaly may only be perceived due to a model-framed illusion \cite{WG2013b}. More generally,  bottom-up risk factors associated with a given instrument can influence the prices,  such as price impact or instantaneous liquidity effects in the presence of investor herding \cite{LM1999, CB2000, S2000}. Hence, price evolution may be driven by both bottom-up and top-down causation, where the latter are described by exogenous factors, leading to  innovations in couplings and feedbacks through the many layers of hierarchy within a financial system: between countries, companies, investment funds, hedging strategies, fund managers, dealers and traders, and bids and asks in the price discovery process.

At different levels, which are characterised by different phenomena, one expects different theories to function.  At some levels in the hierarchy there may be little stability in these effective theories, while stability may be enhanced at other levels. The effective theories that function at the level of the price discovery process, bounded by the market micro-structure at various exchanges on very short time-scales,  may  be more stable than the dominant effective theories which drive prices at the level of investment strategies that operate on much longer time scales. These in turn may be coupled together in complex and interesting ways that are fundamentally nonlinear\cite{WG2013b}.

It is also the case that different theories may be effective at the same level. The number of different plausible theories is a decreasing function of available information, since deeper or broader insight about system workings can eliminate
 theories which become inconsistent under the incorporation of new information.


    In the next section we give an overview of how markets may be mapped into a multilevel causality framework, with consideration of key concepts. In Section III we discuss specific agent-based models to be incorporated in a hierarchy of complexity and in Section IV    we  clarify how risk, information and innovation lead to emergent prices in a concrete multilevel model. In Section V we provide a general discussion of exemplars for actors which drive top-down causation and our conclusion is given in Section VI.





\vspace{-0.2cm}

\section{Towards a hierarchical view of financial markets}

\vspace{-0.1cm}

\subsection{Reductionist perspectives}

\vspace{-0.1cm}

Notions of market equilibrium and price efficiency \cite{K2003,FG2008,K2010,CTSBCHK2012} are fundamental to orthodox models in finance. Equilibrium dynamics in our application relate to how prices are influenced by risk and information and how these are used by various agents in the system. It useful to recall alternative views of how markets can be considered to function in this regard:

\begin{itemize}
\item [EMH] {\it Efficient Market Hypothesis:} \cite{S1965,F1970} Here prices are fully rational, they reflect
{\em fair-value} and are the best predictors of future prices because price changes
can only be due to unpredictable news. This is a very broad aggregative perspective. This is considered to be a normative idealised state that captures salient features of stock market in equilibrium. Weakened forms of this perspective attempt to reconcile this with actual market behaviour.
\item [NMH] {\it Noisy Market Hypothesis:} \cite{Black1986,JS2006} Here there is a meaningful {\em fair-price}
for companies, but speculative noise trading pushes prices away from this fair-value. This is a defense of
the efficient market hypothesis in that it provides a mechanism for prices to deviate from fully rational prices. It is assumed that there is a mean-field theory that can capture the market's salient features after the noise has been averaged away.
\item [AMH] {\it Adaptive Market Hypothesis:} \cite{FL1999,L2004} Here prices do not
necessarily reflect {\em fair-values} . The market is a machine that sets prices. On the one hand, inefficiencies can be substantial, and on the other , patterns in prices may disappear as agents evolve profitable strategies to exploit them. This represents a form of scientific pluralism and is a significant departure from the efficient market view in that there is much more than just noisy deviations from rational prices. There are instead many notions of fair-value and there is no aggregate notion of a fully rational price.
\item [CDEM] {\it Critical Dynamical Equilibrium Market:} \cite{BGPW2004} Here price diffusions are the result of a critical dynamical equilibrium between
persistent effects (liquidity takers which correlate orders) and anti-persistent
effects (liquidity providers which generate mean-reverting forces). No-one can know the {\em fair-value}  reference price, even if it existed. This is the statement that there is no such thing as fair-value and there is no mean-field theory that can capture all salient market features.
\end{itemize}

Clearly these philosophical perspectives on what is attainable offer very different modelling paradigms, ranging from one where accurate prices exist as fundamental information, which can  be discovered, to the assumption that prices are emergent phenomena determined by agents acting on partial and potentially biased information\cite{Black1986, OP2007}.

To consider examples of the type of reductionism, one need only look at Keynesian economics, as described by Shiller \cite{SA2008}: instead of rational bottom-up agents, there are {\em animal spirits} that lead  to aggregate behaviour via bottom-up causality which is greater than that allowed within the efficient market perspective of bottom-up rationality. The approach is still reductionist and places emphasis on the unique attributes of the agents, in particular, their behavioural characteristics make the whole merely as an aggregate of its parts.

In view of consensus that financial markets are more than reductionist \cite{A1995,C2011,FL1999}, we propose a mechanistic combination of causal linkages between levels in a hierarchy of complexity, in tandem with the set of causation classes that act on these levels. Prices in such a system are strongly influenced by both top-down risk factors as well as information variables that constitute bottom-up and top-down state-variables. The prevalence of such drivers are time-varying and depend on market conditions as well as actor interactions that exploit the market models through various types of top-down causation.  Key questions are:  to what extent can an adequate  characterization of financial markets be carried out in a bottom-up manner and which  features of financial markets can only be possible through top-down causation within a hierarchy of complexity?

\vspace{-0.2cm}

\subsection{Levels of complexity}

Modelling financial market dynamics via a large number of interacting agents, who function at various levels within a financial market, is of interest because it is a natural interpretation of real-world interactions.  Relatively simple models can be built to capture essential features in the hierarchy of complexity.

In our approach, asset prices are  key observable quantities of the system and reflect how the system acts on information, as conditioned by the prevailing classes of causation driven by actors who make decisions.

 We use the idea of {\em actors}, which serve as are exemplars of causation types, to describe how effective theories at different levels can be used to account for complexity and adaption in a given financial market.

The following list summarises  a hierarchy of complexity at which effective theories hold, coupled to top-down and bottom-up causation, in a manner consistent with theoretical foundations:

\begin{enumerate}
\item [Level 1] Individual agents generate buy and sell decisions for an asset to generate aggregate price information via an {\em Ising model} of agents on a lattice coupled to an external field.
\item [Level 2] Emergence of aggregate asset prices follows from a {\em Potts model} for asset prices. This provides a framework for bottom-up price emergence from collective interactions in price dynamics which include  top-down influences, modelled by an external field.
\item [Level 3] At this level, we propose an affine  model which describes aggregate asset pricce fluctuations,  driven by net market interpretation of bottom-up  information variables and shared market risk factors. Assuming that additional influences amount to long-term, uncorrelated noise, this level can be represented by a neo-classic arbitrage pricing  theory approach.
\item [Level 4] Prices are impacted by shared, top-down financial and economic factors which  arise  beyond trading activity in financial markets.
\item [Level 5]  Market structure incorporates the regulatory framework prevailing within a market, across markets, as well as more generic market structure, segmentation and market micro-structure. This level influences and determines the nature of arbitrage pricing and the  attainable ranges for prices.
\item [Level 5] Beliefs, meaning and ethics create and influence behaviour in the greater economy. This level of causation level abstracts how an economy may be organised psychologically through agent beliefs, anticipations, expectations, interpretations and ethics \cite{A1995}.
\end{enumerate}

The broad view considered is summarised in Table \ref{tab:levels},  where we represent structures and processes acting across the levels of the hierarchy of complexity to capture the key averaging scales of the system.

\begin{table}
	\begin{tabular}{|l|p{3.3cm}|p{4.2cm}|}
		\hline
			Level & Model & Processes \\
		\hline  \hline
			6 & Meaning and Ethics & \\
		\hline
			5 & Market-structure. & \\
		\hline
			4 & Economic Factors. \par  (Section \ref{s:risk}) & Macro-economic and Financial market generated risk factors. \\ \hline
			3 & Asset price aggregation \par (Section \ref{s:infoandrisk}) & Risk factors, information variables, and noise combine to form aggregate assets prices. \\		\hline
			2 & Asset price emergence  \par (Section \ref{s:potts}) & The collective behaviour of assets generate noise in the presence of different kinds of information.\\		\hline
			1 & Agent interactions \par (Section \ref{s:ising}) & Agents buying and selling an individual asset in the presence of different kinds of information to generate prices. \\		 \hline
	\end{tabular}
	\caption{Hierarchy of complexity given by levels and associated processes. Levels can be coupled both up and down the hierarchy through various feedbacks. Each level has a prevailing effective theory that can be implemented as a model. }
	\label{tab:levels}
\end{table}

\vspace{-0.1cm}
\subsection{Computational implementations}

Agent-based modelling is a well-established discipline within heterodox economics \cite{Follmer1974,AEY2000,Duffy2001,Hommes2006,Tesfatsion2006,LBT2008}. We identify {\em actors } \cite{Hewitt2011,HBS1973}, which serve as exemplars for the types of causation under consideration.   An actor is a computational model, which is used here to  characterise a specific type of top-down causation, and actors incorporate universal features of the level at which they function. Hence, one may differentiate between the concepts of {\em agents} and {\em actors}  as one
differentiates between bottom-up, interacting model components and top-down causation. Actors may explicitly accommodate concurrency and indeterminism and the fact that there is a reduced set of causation classes leads to a reduced set of actors, differentiated by their actions.

Table \ref{tab:actors} describes possible actors functioning across the hierarchy of complexity.

\begin{table}
	\begin{tabular}{|c|p{4cm}|}
		\hline
			Causation Type & Exemplar Actor \\
		\hline  \hline
			TDC1 (section \ref{sec:predict}) & {\bf ``Predictors"}  \\ \hline
			TDC2 (section \ref{sec:profit}) & {\bf ``Profiteers"}  \\ \hline
			TDC3 (section \ref{sec:invest})& {\bf``Investors"}    \\ \hline
			TDC4 (section \ref{sec:trade})& {\bf ``Traders"}     \\ \hline
			TDC5 (section \ref{sec:reg})& {\bf ``Regulators, Rulers"}    \\ \hline
	\end{tabular}
	\caption{Actors serve as exemplars for causations. They are able to act through-out the hierarchy of complexity given in Table \ref{tab:levels}. }
	\label{tab:actors}
\end{table}


The use of an {\em actor model} as a possible computation metaphor to describe exemplars of possible core classes of top-down causation is relevant as a complement to an agent based modelling approach, which is effective for capturing bottom-up interactions\footnote{Top-down causation may be introduced agent based models via  environmental variables, for example, the use of an external field when approximating agents with spin models, or through changing network structure or topology for the domain supporting the spin-models.}.
Financial markets are evolving open-systems, a feature that poses interesting problems for many computational models, for example, a bottom-up agent based modeling framework and actor models offer a relatively simple capacity to cope naturally with {\it indeterminism} and {\it concurrency}.

Three core features characterise an actor \cite{HBS1973, Hewitt2011}: (i) the ability to process information, (ii) the ability to store information, and (iii) the ability to communicate. When combined with it functional axioms: (A1)  actors can create more actors,  (A2)  actors can send messages  and (A3)  actors decide what to do with the next message.
Although messages are delivered at most once, without duplication, the actor model is a many-to-many model in which the use of a metaphor of identity of an actor and an actor's address  can enable a single actor to be associated with one or more addresses and enable one address to be associated with one or more actors by their identity. This creates the potential for a very fluid dynamic between actors that has the possibility of high-concurrency.

The next feature of interest is the actor model's ability to generate and handle indeterminism \cite{Hewitt2011}. Indeterminism is a foundational feature of any system that is truly concurrent, because  there is nothing to ensure or tag the order in which messages need to be managed. Hence, the system has to order things by itself and in the actor model, indeterminism is handled naturally using the concept of arbiters, which take simultaneous inputs and generate sequential outputs.

A consequence of the actor model for computation is that a global consensus is not possible. In the language of spin-models, this can be understood in terms of the phenomenon that there is no mean-field theory. However, this feature encodes a deep form of pluralism within a given system, where local arbitration is possible but one local view can really be quite different from another. Such local positions can be inconsistent but still be functionally important to the viability of the system itself.
Thus, the actor model can be used to encode top-down causation, as distinct from the bottom-up causation  which is natural to the agent based model perspective.

Conceptual issues relating to {\it cohesion} and {\it coupling} \cite{SMC1974}  are at the heart of approaches to fragility and robustness in complex software systems.  Here, we recall that a goal of software development is to build a system with with low-coupling and high-cohesion in order to create a system which is both robust and fit-for-purpose \cite{SMC1974}. Cohesion relates to the degree to which elements in a system belong together in terms of their function or responsibilities. This is differs from procedural coupling within a system, which is concerned with the degree to which processes and components of the system depend or rely on each other.

Since similar features are relevant for the modern financial system, the actor model approach incorporates economically meaningful elements: from a regulatory perspective, one would be aim to avoid tight-coupling (strong internal dependencies) in a financial market, particularly when the system may have low-cohesion and elements of the system do not belong  or connect.


\vspace{-0.4cm}

\subsection{Emergence}

  Aristotle claimed that \cite{A}
  \begin{quote} The whole is something over and above its parts, and not just the sum of them all...\end{quote}

  Emergence refers to the concept that novel and coherent, distinguishable structures or patterns can arise during the process of self-organization in complex systems \cite{A, P1926, A1972, G1999,C2002}.  Our approach is narrowly confined to how random interactions lead to some form of weak emergence due to feedbacks \cite{A1972, G1999,C2002}.


Even though noise provides additional degree's of freedom that are important for adaption\cite{Black1986}, it has been shown that by adding uncorrelated noise to oscillators one can induced explosive synchronization, i.e., very sharp phase transitions \cite{K1984,SA2014}. Barab\'{a}si and Albert show that for sufficiently heterogeneous network topologies, simple correlations can induce explosive phase transitions \cite{BA1999}.

 Thus, the role of noise within the hierarchy of complexity needs to be considered with care, since the related capacity for creating opportunities for innovation can provide pathways to tight-coupling and lead to fragility or even catastrophic failure.


In order to understand the nature of the couplings across the hierarchies,  phenomenological features within a given level must be identified to allow quantification of coupling strengths. A mechanistic approach towards monitoring may be futile within a highly adaptive complex system, such as a financial market, since it is only in hindsight that tightly-coupled paths of causation may become apparent. It may be more feasible to identify observable universal features of various top-down causation types at different levels in the hierarchy of complexity, while discerning how bottom-up features may trigger events based on tightly-coupled top-down and bottom-up causation features.

\vspace{-0.2cm}
\section{Agent-based perspectives}

\subsection{An Ising market model for a single stock} \label{s:ising}

Key strategies applied in agent based models to generate prices for {\em a single asset} are\cite{LM1999,CB2000,F2002}:
\begin{itemize}
\item {\em do what you neighbours do}, which captures herd behaviour of noise traders, for example as in  the Lux-Marchesi model \cite{LM1999}, and
\item {\em do what the minority does}, which captures the behaviour of traders with opinion about the values of trade-able asset.
\end{itemize}

These conflicting interactions have been combined into a simple spin model by Bornholdt \cite{B2001} where the standard nearest neighbour interactions capture the neighbour interactions, while the coupling to the global observables is captured via the global magnetization of the system. In Bornholdt model the local field $h_i(t)$, which is used to describe net demand or net price increment,  takes on the form:
\begin{eqnarray}
h_i(t) = \sum_{j=1}^N J_{ij} s_j - \alpha C_i(t) \frac{1}{N} \sum_{j=1}^N s_j(t).
\end{eqnarray}
Here the model has ${i=1,\ldots,N}$ spins with orientation $s_i(t) = \pm 1$. The first term is the local Ising Hamiltonian with nearest-neighbour interactions $J_{ij}=J$ and $J_{ij}=0$ for all other interactions. This induces order based on a given scale of
interaction. The second term is the global coupling where a given strategy employed by the $i$-th agent is captured by the spins $C_i(t)$.

The local field determines the dynamics of the spins according to
\begin{eqnarray}
s_i(t+1) &=& +1 ~\mathrm{with}~ p_+={1 \over{1+e^{-2\beta h_i(t)}}}, \\
s_i(t+1) &=& -1 ~\mathrm{with}~ p_-=1-p.
\end{eqnarray}
The simple case described by Bornholdt sets $C_i(t) =+1$. This creates a force that aligns the minority of spins in the system. These types of traders can be considered as fundamentalists as each agents tries to align relative to fundamental value and as such are contrarian in nature; this would suppress large fluctuations in the system.

The case with $C_i(t)=-1$ would cause alignment with the majority of agents in the system and as such can be considered a proxy for chartists where agents follow the broad consensus views and as such herd with the global behaviour; this cause global alignment of the system and as such allows large fluctuations in the system.

For a fixed ratio of chartist and fundamentalist in the system the chartist would dominate. Therefore, an interaction term allowing for transitions between chartists and fundamentalist was introduced \cite{B2001}:
\begin{eqnarray}
C_i(t+1) = -C_i(t) ~\mathrm{if}~ \alpha s_i(t) C_i(t) \sum_{j=1}^N s_j(t) <0.
\end{eqnarray}
Here the idea is that an agent in the majority will choose the strategy $C_i(t)=+1$ while the agent in the minority will choose the strategy $C_i(t)=-1$ so that each agent chooses a risky strategy for the prospect of higher returns, but at a cost.

Agents will try to anticipate losses given a participation threshold and attempt to switch to that asset which will become the next majority or minority asset depending on the trading strategy. This model takes on a simplified local field when strategy adjustments are considered instantaneous:
\begin{eqnarray}
h_i(t) = \sum_{j=1}^N J_{ij} s_i - \alpha s_i \left| {\frac{1}{N}\sum_{j=1}^N s_j}, \right| \label{eqn:bornholdt1}
\end{eqnarray}
with a global coupling constant $\alpha>0$. The first term aligns spins, while the second terms causes the spins to change sign if the global magnetization become large.

To understand the interpretation of the model in the context of a financial market one can interpret the spins states as demand for an asset. The magnetization term varies with net demand and represents a function of aggregate price change \cite{CB2000,F2002}:
\begin{eqnarray}
X(t) = \frac{1}{N} \sum_{j=1}^N s_j. \label{eqn:bornholdt2}
\end{eqnarray}

Price fluctuations are the observable property of the system and a key success of this model is that it exhibits intermittency  and  volatility clustering, as well as large jumps in returns, without necessitating fine tuning. One could introduce an additional parameter to index each stock and in a more generic framework one should also consider the excess demand in more generality, as a function of price and market depth in order to solve a market clearing equation where the aggregate magnetization is zero across the entire system.

For the descriptive purposes used here, we assume linearity in demand function with a constant market depth to allow us to view the magnetization as a measure of the induced price changes {\footnote{In a linear model \cite{F2002,F2003} for price change we can consider a price $p(t)$ such that $\frac{d p(t)}{dt} = \frac{1}{\lambda} M(t) p(t)$ for magnetization $M(t)$ define as the aggregate sum of spin states in the system as a representation of excess demand. Under discretization this implies that $p(t+1) = p(t)  e^{c M(t)}$ to allow one to compute the price change : $X(t) = \ln(p(t)) - \ln(p(t-1)) = c M(t-1)$.} }.


We now move to a Potts model description of the price dynamics of collections of stocks to introduce a second level in our hierarchy of complexity.

\subsection{A Potts model market for interacting stocks} \label{s:potts}

Stocks whose prices evolve under correlated dynamics may be group into co-evolving collections or clusters. A key idea underlying such a group based stock pricing model is that we seek to reflect two distinct features:
\begin{itemize}
\item {\em a group of stocks have something in common}, i.e. the shared behaviour within a groups of stocks can be described by modelling a  the systematic component.
\item {\em a stock's price can include unique information}, i.e. the model should include non-systematic behaviour which is unique to a particular stock.
\end{itemize}

One can apply the super-paramagnetic ordering of a $q-$state Potts model directly for cluster identification\cite{BWD1996}. In a Potts model market,  each stock may only belong to one of  $q-$states \cite{BWD1996, KKM2000,GM2001} and each state can be represented by a cluster of similar stocks. Cluster membership is indicative of some commonality among the cluster members and each stock has a component driven by state dynamics  and a component influenced by stock specific noise. In addition,  there may be global couplings that influence all the stocks, represented by the external field to model a market mode.

In this approach, the cost function can be interpreted as a Hamiltonian whose low energy states correspond to cluster configurations that are most compatible with the data sample. Structures are  identified  with configurations, denoted ${\cal S} = \{ {s_i} \}_{i=1}^N,$ where the $s_i$   denotes the equivalence class  or cluster to which the $i$-th object belongs. The Hamiltonian takes on the form:
\begin{eqnarray}H  = - \sum_{s_i,s_j \in S} J_{ij} \delta(s_i,s_j) - \frac{1}{\beta} \sum_i k_i s_i.
\end{eqnarray}
Here, the spins $s_i$ range over $q$-states and can be interpreted as  spins in the Potts model  \cite{BWD1996, KKM2000}, with  $k_i$ serving to tune external influences. The first term represents common internal influences and the second term represents external influences.  One can interpret the coupling parameters $J_{ij}$ as being  functions of correlation coefficients \cite{KKM2000,GM2001} in order to fix  a distance function that is decreasing with distance between objects. If all the spins are related in this way, then each pair of spins is connected by some non-vanishing coupling $J_{ij}=J_{ij}(c_{ij})$.

 The case where there is only one cluster can be thought of as a ground state. On the other hand, if the system becomes more excited, then  it could break up into additional clusters. Each cluster would have specific Potts magnetizations, even though the nett magnetization may be zero. Generally, correlations may be both a function of time and a temperature factor in order to encode both the evolution of clusters as well as the hierarchy of clusters modelled as a function of temperature.

We seek the  the lowest energy state that fits the data. In order to parameterize the model efficiently one can choose to make the Noh ansatz \cite{Noh2000} and use this to develop a maximum-likelihood approach \cite{GM2001} rather than explicitly solving the Potts Hamiltonian numerically \cite{BWD1996,
KKM2000}. In this investigation, we ignore the second term when fitting data because we include shared factors directly in later sections when we discuss information and risk and the influence of these on price changes.

To interpret the system observables as being the cluster configurations which evolve through time, the  price increments associated with the $i$-th stock may be given by:
\begin{eqnarray}
X_i(t) = g_{s_i} \eta_{s_i} + \sqrt{1-g_{s_i}^2} \epsilon_i ,\label{eqn:giada-marsili}
\end{eqnarray}
where the cluster related influences are driven by $\eta_{s_i}$ and the stock unique effects by $\epsilon_i$. Both parameters are assumed to be mutually independent Gaussian random variables, with unit variance and zero mean \cite{foot1}. The relative contribution of the group component to returns is controlled by the intra-cluster coupling parameter $g_{s_i}$.

This model encodes the idea that stocks that have something in common are in the same cluster but also that stock membership in clusters is mutually exclusive and that intra-cluster correlations are positive.

\vspace{-0.3cm}

\section{Hierarchies, risk and emergence: a multilevel model} \label{s:risk}

We exhibit how risk and information may be incorporated in a level within a hierarchical system such that the basic features of arbitrage pricing theory are recovered as a level within the hierarchy of causation. The level is described by a family of linear risk and information models, where a rationale for the family of models is reviewed in detail in Wilcox and Gebbie\cite{WG2013b}.
We note that more advanced stochastic modelling may be used to analyse {\em derivative securities} in incomplete markets under assumptions of semi-martingale dynamics for underlying processes\cite{FS1991,KQ1995,OS2014}.

First, we describe three simplified assumptions which characterise a simple risk-based pricing model:
\begin{itemize}
\item {\em Agents are only interested in means and covariances} for a view of the underlying distribution of asset prices and their relationship to underlying risks.
\item {\em Expected returns in excess of a risk-free asset only depend on their relationship to shared risk factors} to reflect the idea that non-systematic behaviour which is specific to a particular stock is short-lived.
\item {\em There is a unique risk-free asset} for tractable risk-neutral modelling perspectives.
\end{itemize}

Next we review a standard model for shared risks driving asset prices, using the idea of no-arbitrage pricing. We then perturb this view by directly including information variables.

\vspace{-0.2cm}
\subsection{Shared risk} \label{ss:risk}

The Fama and French \cite{FF1992,FF1996} model for accounting for temporal risks associated with unanticipated returns,
\begin{eqnarray}
 r_{i,t+1} - \mathrm{E}_t[r_{i,t+1}],
\end{eqnarray}
has served as a standard approach to using arbitrage pricing theory. More generally, the times-series of excess returns may be modelled as follows:
\begin{eqnarray}
r_{i,t+1} - \mathrm{E}_t [r_{i,t+1}] = \sum_{p=1}^P \beta_{i,p,t} \left[{ r_{p,t+1} - \mathrm{E}_t[r_{p,t+1}]}\right ] + \epsilon_{i,t+1}, \label{unantret}
\end{eqnarray}
where $r_{i,t+1}$ is the excess return of the $i$-th stock and $r_{p,t+1}$ are the excess returns on the factor-mimicking portfolios
that proxy potential shared risks, here for the $p$-th risk factor. This shared risk could be of a statistical nature, for example the eigenmodes of the correlation
matrices of asset price returns \cite{WG2007}, or they could be of a fundamental nature. The $\epsilon_{i,t}$ are zero-mean noise contributions which are uncorrelated with the risk factors, but may be correlated across assets.

Now suppose the following serves as a model for the conditional expected returns:
\begin{eqnarray}
\mathrm{E}_t[r_{i,t+1}] = \alpha_{i,t} + \sum_{p=1}^P \beta_{i,p,t} \mathrm{E}_t[r_{p,t+1}]. \label{sharedrisk}
\end{eqnarray}
If this is substituted into Eqn. (\ref{unantret}) and $r_{i,t+1}$ is replaced with $\tilde r_{i,t+1}$,  to indicate that the quantity is now the measured
return (where $\hat r$ is used to denote estimated returns), then we obtain a conventional pricing model used in financial economics \cite{FF1996}:
\begin{eqnarray}
\tilde r_{i,t+1} = \alpha_{i,t} + \sum_{p=1}^P \beta_{i,p,t} \tilde r_{p,t+1} + \epsilon_{i,t+1}.
\end{eqnarray}
Early convention wisdom was argued that if risk factors in the model explain returns,  then the {\em alpha}  term, $\alpha_{i,t}$, should be vanishingly small. The
typical approach is to use time-series regression to estimate a static {\em beta}  for a sample
 period, {\it i.e.} $\mathrm{E}_t[\beta_{i,p,t}] \sim \hat\beta_{i,p} = \hat \Sigma_{i,p} / {\hat\sigma_p}^2$ based on the
sample estimated covariance matrix.

A limitation of this model is that there may be variables that influence asset prices in practice which  have little to do with shared risks. We refer to these variables as  shared information variables.


\vspace{-0.1cm}
\subsection{Shared information} \label{ss:info}

Haugen and Baker (1996)\cite{HB1996} provide a  useful approach in financial economics for incorporating shared information variables.  Their model has a similar form to that for shared risks as in Eqn. (\ref{sharedrisk}), but expected returns $\mathrm{E}_t[r_{i,t+1}]$ are now estimated in terms of lagged attributes. The latter are typically company or asset specific attributes which are bottom-up information variables, denoted by $\tilde \theta_{i,m,t}$, for the $m$-th source of information relevant to the $i$-th stock at time $t$. The cross-sectional model specification takes on a linear form as follows:
\begin{eqnarray}
\tilde r_{i,t+1} = \alpha_{t+1} +\sum_{m=1}^M \delta_{m,t+1} \tilde \theta_{m,i,t} + \epsilon_{i,t+1}
\end{eqnarray}
The conditional expected returns for the $i$-th stock can be estimated by:
\begin{eqnarray}
\mathrm{E}_t[r_{i,t+1}] = \mathrm{E}_t[\alpha_{t+1}] + \sum_{m=1}^M \mathrm{E}_t[\delta_{m,t+1}] \tilde\theta_{i,m,t} \sim \alpha_{i,t}, \label{HBrisk}
\end{eqnarray}
where the expected pay-off, $\mathrm{E}_t[\delta_{m,t+1}]$, to the lagged information variable can be estimated by a variety of means to compute the expected conditional returns.

This model is not consistent with the Fama-French \cite{FF1996}
or arbitrage pricing theory models. However, it includes features which support avenues for top-down and bottom-up feedbacks. In addition to the model's practical usefulness, for small deviations from the arbitrage pricing theory framework, one can combine information variables with risk variables and retain a risk interpretation for asset pricing in the long-time limit.

\vspace{-0.1cm}
\subsection{Information, risk and causation}\label{s:infoandrisk}

Ferson and Harvey \cite{FH1999} combine time-dependent risk factors in Eqn (\ref{sharedrisk}), the $ \mathrm{E}_t[r_{p,t+1}]$, with lagged asset specific information variables by conditioning the $\beta$ coefficients with time-dependence as follows: $\beta_{i,p,t}= \beta_{i,p,t}(\theta_{i,m,t}) := \left[ { {b^0}_{i,p} + \sum_{m=1}^M {b^1}_{m,p} \theta_{i,m,t}}\right]$. This is analogous to the manner in which the time-$t$ expectation of $\delta_{m,t+1}$ is computed from $\theta_{i,m,t}$ for Eqn.  (\ref{HBrisk}). Hence, the following model specification for the conditional expected returns is obtained:

\begin{eqnarray}
\mathrm{E}_t[r_{i,t+1}] = \alpha_{i,t} + \sum_{p=1}^P \beta_{i,p,t}(\theta_{i,m,t})\mathrm{E}_t[r_{p,t+1}]. \label{FSrisk}
\end{eqnarray}

Thus, by  substitution, the time $t+1$ measurements can be obtained by expanding Eqn. (\ref{unantret}) to obtain the following explanatory model for returns:
\begin{eqnarray}
\tilde r_{i,t+1} &=& \left\{ {\alpha_{i,t} + \sum_{\substack{p=1 \\ m=1}}^{P,M} {b^1}_{m,p} {\tilde \theta}_{i,m,t} {\tilde r}_{p,t+1}} \right \} \nonumber \\ &&+ \sum_{p=1}^P {b^0}_{i,p} {\tilde r}_{p,t+1} + \epsilon_{i,t+1},
\end{eqnarray}
where writing in the original Haugen and Baker variables would give the following representation:
\begin{eqnarray}
\tilde r_{i,t+1} &=& \left \{ {\alpha_{i,t} + \sum_{m=1}^M \delta_{m,t+1} \tilde \theta_{i,m,t} } \right \}\nonumber \\ & &  + \sum_{p=1}^P {b^0}_{i,p} \tilde r_{p,t+1} + \epsilon_{i,t+1}.
\end{eqnarray}

Since the estimated coefficients in Eqns  (\ref{HBrisk}) and (\ref{FSrisk}) are obtained by cross-sectional regression, the model is nonlinear and has been  supported by empirical evidence\cite{WG2013b}. The combined model has two important model features. Firstly,   shared risk factors include time-dependence on the information variables and secondly,
the model conforms with arbitrage pricing theory as long as the non-vanishing bias terms, the so-called $\alpha$ terms, only have short-term departures from being negligibly small, and $\beta$'s are allowed to be time-dependent.

 Using the above insight, we propose the following factorisation of the right-hand-side of Eqn. (\ref{HBrisk}) for a unifying framework to incorporate top-down and bottom-up economic information. The model is built on the two sources of causality:
 \begin{itemize}
 \item generic information variables, denoted $Z_{k,t}$ for the $k$-th source of top-down information, where $Z_{k,t}$ referred specifically to macro-economic information in \cite{FH1999}, and
\item  company specific variables, $\theta_{i,m,t}$ as in Haugen and Baker \cite{HB1996}, which serve as bottom-up information variables , now in a single unified framework.
\end{itemize}

Following the approaches of Haugen and Baker \cite{HB1996} and Ferson and Harvey \cite{FH1999}  by effectively conditioning the  coefficients of both the top-down market information and bottom-up stock-specific information terms using cross-sectional regression on previous-time top-down and bottom-up information. This yields this following representation for coefficients:
\begin{eqnarray}
\alpha_{i,t} &=& {\alpha^0}_i + \sum_{k=1}^K {\alpha_{i,k} \tilde Z_{k,t}}, \nonumber \\
\beta_{i,p,t} &=& \sum_{p=1}^P \left[ {{b^0}_{i,p} + \sum_{k=1}^K {b^2}_{i,k,p} \tilde Z_{k,t} + \sum_{m=1}^M {b^1}_{m,p} \tilde \theta_{i,m,t}} \right]. \nonumber
\end{eqnarray}

Putting it all together we obtain:
\begin{eqnarray}
{\tilde r}_{i,t+1} &=& \left \{ { {\alpha^0}_i + \sum_{k=1}^K {\alpha^1}_{i,k} {\tilde Z}_{k,t} +\sum_{\substack{p=1 \\ k=1}}^{P,K} {b^2}_{i,k,p} {\tilde Z}_{k,t} {\tilde r}_{p,t+1}} \right \} \nonumber \\
&& +\sum_{\substack{p=1 \\ m=1}}^{P,M} {b^1}_{m,p} {\tilde \theta }_{i,m,t} {\tilde r}_{p,t+1} \nonumber \\
&&+ \sum_{p=1}^P {b^0}_{i,p} {\tilde r}_{p,t+1} + \epsilon_{i,t+1}.
\end{eqnarray}

This gives a theory that is driven by the returns of shared risk as proxied by the factor-mimicking portfolios $r_{p,t}$ but conditioned by top-down information variables $Z_{k,t}$ (for example slowly changing economic features) and bottom-up information variables $\theta_{i,m,t}$ (for example asset specific features) in the presence of potentially correlated noise $\epsilon_{i,t}$.

\begin{table}
	\begin{tabular}{|p{2.2cm}|p{3.8cm}|c|p{0.8cm}|}
		\hline
		Variable Type & Description & Symbol & Level \\
		\hline  \hline
			 Risk factors & The factor mimicking portfolio for the $p$-th risk factor at time $t+1$ from state variables $Z_{k,t}$ and $\theta_{m,t}$. & $r_{p,t+1}$  & 5-4 \\ \hline
			Top-down information &  The $k$-th shared state variable at time t.  &  $Z_{k,t}$ & 5-3 \\ \hline
			Bottom-up information & The $m$-th state variable for the $i$-th asset at time t.  &  $\theta_{i,m,t}$ & 5-2\\ \hline
			Cluster specific noise & The noise term for $s_i$-th clusters at time t. & $\eta_{s_i,t+1}$  & 5-2\\ \hline
		          Asset specific noise &  The noise term for i-th asset at time t.  & $\epsilon_{i,t}$  & 5-1 \\ \hline
	\end{tabular}
	\caption{The key variables in the model as driven by noise formation using spin models and the top-down and bottom-up information variables that describe both top-down risk factors as well as there time-dependence. Actors are both able to influence the state-variables as well as select for them as part of the various causation classes. }
	\label{tab:variables}
\end{table}

This  particular framework remains consistent with arbitrage pricing theory \cite{WG2013b} and illustrates that, even within arbitrage pricing theory,  departures from equilibrium are permitted through either top-down or bottom-up causation between different levels in the hierarchy of complexity.

The finally modification we make to our model is to modify the noise term to accommodate emergence and complexity via innovations in coupling in the agent based model. The key variables for the combined model are summarised in Table \ref{tab:variables}.

\vspace{-0.2cm}
\subsection{Emergence} \label{s:emergence}

In our model, we address scenarios where stocks group together randomly on short-time scales due to liquidity, or lack thereof, asynchronicity of trading or market micro-structure effects, even if econometric similarities and shared risk between assets still drive stock clustering over longer horizons.  Such incidence could introduce non-trivial lead-lag and feedback effects.  To model this idea, we apply the group model of Noh \cite{Noh2000} (following the approach of Giada and Marsili\cite{GM2001}). This enables us to consider a cluster noise term $\eta_{s_i}$ for unanticipated price changes (due to the random fission and fusion of mutually exclusive groups of assets) into order to describe unanticipated returns in Eqn. (\ref{unantret}) as follows:
\begin{eqnarray}
r_{i,t+1} &-& \mathrm{E}_t [r_{i,t+1}] = \alpha_{i,t}  \nonumber \\
&& + \sum_{p=1}^P \beta_{i,p,t} \left[{ r_{p,t+1} - \mathrm{E}_t[r_{p,t+1}]}\right ] \nonumber \\
&&   + g_{s_i} \eta_{{s_i},t+1} + \sqrt{1-{g_{s_i}}^2} \epsilon_{i,t}. \label{emergence}
\end{eqnarray}

Here, as before, $s_i$ denote Pott spins of the cluster that the $i$-th stock currently resides in and  $n_s$  denote  finite, but variable number of such clusters in the market at any given time. The binding strength of the $s_i$-th cluster components is given by $g_{s_i}$, the uncorrelated cluster specific noise is described by $\eta_{{s_i},t}$ and the stock specific uncorrelated noise which is independent of random cluster emergence is denote by $\epsilon_{i,t}$ for the $i$-th stock.

The final step in our model specification is to pass to conditional expected returns in Eqn. (\ref{emergence}) above, analogous to the manner that expected future returns were described in Eqn. (\ref{unantret}).

We have demonstrated a model which couples together different parts of a hierarchy of complexity and which permits random innovation leading to innovation, into a single pricing equation. Before concluding, we discuss specific examples of top-down causation and possible roles.

\vspace{-0.3cm}
\subsection{Block variables and hierarchies of complexity}

It is well understood that money serves the purpose of smoothing over mismatches in supply and demand at the level of direct bartering to facilitate transactions. More generally, the ability to transact instantaneously can be measured by estimating \textit{liquidity}. Black\cite{Black1986}, Ellis \cite{GFRE2008} and Kirman and Helbing \cite{HK2013} observe that noise is a fundamental feature of complex adaptive systems because it adds slack by allowing additional degrees of freedom to the set of options available in a given systems.  We consider how noise modelling in complex systems facilitates features such as  the ability or incentive to adapt through selection. Typical responses of the top-down causation classes to noise generated in complex adaptive systems are given in Table \ref{tab:tdc-noise}.

One approach to modeling a hierarchy of complexity within a spin system is by using block variables, following the renormalisation group approach  \cite{K1966,W1979,S2000,Haldane2009,Gorban2006,CB2010}. The rationale  is that larger block domains can be representative of higher complexity structures\footnote{Renormalisation group analysis is a mathematical tool invented for the analysis of critical phenomena in systems which are characterised by structures on many different scales.  These systems exhibit power law relationships between the system observables and control parameters }\cite{W1979,S2000} that arise in critical systems with structures on different scales.


Such systems may typically exhibit one of three broad regimes: (i) the regime where noise effects are minimal and ordering across the system is maximal, (ii) the regime when noise effects dominate the system and ordering effects are minimal, and (iii) the critical regime where changes in the scale of the system do not change the behaviour or properties of the system as the system balances between order and possibility of saturation by noise. 

\begin{table}
\begin{tabular}{|p{3.3cm}|p{4.9cm}|}
\hline
Causation Type & Typical Response to Noise \\
\hline \hline
TDC1 (section \ref{sec:predict}) & Noise is a hinderance and undermines the adopted model. \\ \hline
TDC2 (section \ref{sec:profit})  & Noise effects are reduced because a fixed goal is aimed at averaging away noise. \\ \hline
TDC3 (section \ref{sec:invest})  & Noise facilitates or is exploited in the long-term through adaption. \\ \hline
TDC4 (section \ref{sec:trade})   & Noise both facilitates and is created for adaption under selection. \\ \hline
TDC5 (section \ref{sec:reg})     & Noise both facilitates and is created and system goals and selection criterion are adapted to exploit accordingly. \\ \hline
\end{tabular}
\caption{The table summarises some typical responses of the causation classes to noise and the feedbacks associated with noise.}
\label{tab:tdc-noise}
\end{table}

The first limitation of a  block variable view is that stocks are not seen as emergent properties of the system. Secondly, we have already posited that a complex financial  market cannot be determined fully  by using only bottom-up averaging implicit in such a block variable approach. Thus, representing a stock's price as magnetization of a particular block may not be sufficiently realistic. Nevertheless,   as an approximation towards modeling a hierarchy of complexity, we argue that the analogies are useful.


 We have provided particular, ad hoc representation and approximation of the processes driving price increments via emergence, whereby the stock itself, in the sense of the existence of a company, is not an emergent property of the system, but the price of a stock is.

 More generally, one may model companies in a complex exchange economy as represented by stock prices, which arise as emergent properties of the system, with top-down and bottom-up causation features in agreement with the more general situation. This would entail a progression in models from many agents trading a single stock, to agents trading groups
 of stocks, and so on, up to the point where one has many agents trading the market, and groups of markets within a global system.
As in models for physical systems, the insight of averaging to identify dominant dynamics in each level can help build an effective model that agrees with the observed  phenomenology at each scale.

In the scheme described in this paper, the top-down risk and information variables, ($r_{p,t}$ and $Z_{k,t})$, respectively, can be view as external fields in the effective models, while the bottom-up information variables $\theta_{i,m,t}$ serve to model structural features of the system itself. This offers a simplified, but quantitative perspective of the interaction complexities which can and do occur in financial markets.


Viewing the system through a lens of such a hierarchy of complexity can lead to insights at various levels. At a given level, one may consider attempting to regulate interactions, for example to be constrained by productivity or transformative goals \cite{A2004}. However, this may be {\em ineffective} if either there is significant adaptation at time-scales which are shorter than those of regulatory response or emergence is susceptible to noise effects or random bottom-up interactions, possibly triggered by regulation itself.

 Top-down control can also induce noise effects in  lower levels, which may amplify or induce bottom-up effects that may swamp any attempts of control the system. In such cases,
 impacts on one set of system variables $Z_{k,t}$ and $\theta_{i,m,t}$ may induce new states, for example of shared risks, and modify evolution dynamics, as reflected in changes in other connected variables.

When viewing the system from lower levels in a hierarchy of complexity,  one may well become increasingly convinced of a critical dynamical equilibrium view of the markets \cite{}. However, an efficient market view may seem valid when viewing the same system through the lens of pricing models with time-scales which  allow  averaging away of noise effects.
Neither view may be entirely wrong. If there is indeed a hierarchy of complexity within the system,  there may be times at which both can be right or wrong.
Hence, a hierarchy of complexity can be consistent with the realities of scientific pluralism inherent in many complex adaptive systems.

Thus, even the choice of parameters in pricing models and how models are used becomes related to the type of causation models adopted by agents in the system.
In the next section we discuss as a selection of examples for classes of actors in models of financial markets.

\section{Top-down causation  actors} \label{s:topdown}


We discuss the causality classes and the actors used to represent them, as described in Table \ref{tab:actors}, with interest in how they respond to noise, as described in Table \ref{tab:tdc-noise}, and respond to and influence the information variables $Z_{k,t}$ and $\theta_{i,m,t}$, that in turn influence the risk factors $r_{p,t}$, as described in Table \ref{tab:variables}. The five types of actors reviewed are discussed with more general examples in Ellis \cite{GFRE2008}, $\S$4.2. Our discussion is by no means comprehensive and further examples may be expanded\cite{F2002}.

\subsection{Algorithmic top-down causation} \label{sec:predict}

 This view of causation in financial markets is essentially a deterministic, dynamical systems view. This can serve as an interpretation of financial markets whereby markets are understood through assets prices, representing information in the market, which can be described by a dynamical system model.  This is the ideal encapsulated in the Laplacian vision \cite{H2010}:

\begin{quote}
We ought to regard the present state of the universe as the effect of its antecedent state and as the cause of the state that is to follow. An intelligence knowing all the forces acting in nature at a given instant, as well as the momentary positions of all things in the universe, would be able to comprehend in one single formula the motions of the largest bodies as well as the lightest atoms in the world, provided that its intellect were sufficiently powerful to subject all data to analysis; to it nothing would be uncertain, the future as well as the past would be present to its eyes. The perfection that the human mind has been able to give to astronomy affords but a feeble outline of such an intelligence. [1814]
\end{quote}


Here boundary and initial conditions of variables uniquely determine the outcome for the effective dynamics at the level in hierarchy where it is being applied.
This implies  that higher levels in the hierarchy can drive broad macro-economic behavior, for example: at the highest level there could exist some set of differential equations that describe the behavior of adjustable quantities, such as interest rates, and how they impact measurable quantities such as gross domestic product, aggregate consumption.

The literature on the {\em Lucas critique} \cite{S1996} addresses limitations of this approach. Nevertheless, from a completely ad hoc perspective, a dynamical systems model may offer a best approximation to relationships at a particular level in a complex hierarchy.

{\bf Example: Predictors: }
This system actor views causation in terms of uniquely determined outcomes, based on known boundary and initial conditions. Predictors may be successful when mechanistic dependencies in economic realities  become pervasive or dominant.

An example of a predictive-based argument since the Global Financial Crises (2007+) is the bipolar {\em Risk-On/Risk-Off} description for preferences\cite{McC2012,Corcoran2013}, whereby investors shift to higher risk portfolios when global assessment of riskiness is established to be low and shift to low risk portfolios when global riskiness is considered to be high. Mathematically, a simple approximation of the dynamics can be described by a Lotka-Volterra (or predator-prey) model.
The excess-liquidity due to quantitative easing and  the prevalence and ease of trading in exchange traded funds and currencies,  combined with low interest rates and the increase use of automation, provided a basis for the risk-on/risk-off analogy for analysing large capital flows in the global arena.

In our Ising-Potts hierarchy, top down causation is filtered down to the rest of the market  through all the shared risk factors, $r_{p,t}$ and the top-down information variables, $Z_{k,t}$, which dominate bottom-up information variables, $\theta_{i,m,t}$. At higher levels, bottom-up variables are effectively noise terms. Nevertheless, the behaviour of the traders in a lower levels can still become driven by correlations across assets, based on perceived global riskiness. Thus, risk-on/risk-off transitions can have amplified effects.

\subsection{Top-down causation via non-adaptive information control} \label{sec:profit}

In this class of causation, in-built and unchanging goals determine the outcome. In general, a goal refers to any targeted outcome. In capitalist economics, the  primary goal of maximization of individual and corporate profit is one of the most pervasive,  unchanging goals in financial systems, second perhaps only to the objective of attaining a living wage. As a result, the profit motive  is encoded in a variety of structural features in the modern financial systems, including  remuneration structures, the tax regime, the legal framework for the flow of capital and the flow of labour, legal structures deployed to manage corporations.

{\bf Example: Profiteers:}
Since most other capitalist goals are subsumed by quests for profit,
we refer to the dominant class of  actors operating in this paradigm as the Profiteers.
While the use of goals may damp out the effects of randomness as the system is driven towards  pre-selected goals\cite{Black1986}\footnote{TG thanks GFR Ellis for interesting discussions on goal oriented control in the presence of noise.},  top-down causation of the profiteers can  also lead to necessary conditions for financial crises. The history of finance is littered with examples of corporate malfeasance that fit into this category of top-down causation, from WorldCom and Enron, and through to the various banking scandals resulting in billions of dollars worth of litigation \cite{AR1993}

Particularly striking is the ease with-which collusion in financial systems may occur when the profit motive dominates decision making.
This can occur when bottom-up information variables, $\theta_{i,m,t}$, concerning individual corporate credit ratings, debt cover or book-values, etc, which initially drive the system,
 become manipulated with disinformation, but are nevertheless presented in a sanitized manner.
Through agent responses, this can in turn influence the behaviour of shared-risks in the market which arise out of  aggregating the information in $\theta_{i,m,t}$ and $Z_{k,t}$.

\subsection{Top-down causation via adaptive selection} \label{sec:invest}

Fixed,  high-level goals guide outcomes of stock selections in this class of causation.  This paradigm is geared towards optimal deployment of assets via maximization of an appropriate utility function (which acts as a fixed meta goal), to allow risk-return decisions to be made in the face of uncertainty. Such decisions are more general than profit maximization. More specifically, in Markowitz optimization for example, which is equivalent to maximising quadratic utility under assumptions of Gaussian increments, investors choose to maximize wealth using the means and covariances.

{\bf Example: Investors:} The influential actors  in this class may be referred to as Investors. To adopt a top-down causation model with adaptive selections, investors require markets to be sufficiently efficient, with departures from equilibrium that are short-lived, uncorrelated noise.
 Such investors have an idealised sense of being able to diversify risk as a meta goal and, hence, the idea of being able to optimally deploy capital. Unpredictability enters the dynamics due to  randomness. In general, high level selection criteria have the advantage of exploiting randomness in an ensemble of options\footnote{TG thanks GFR Ellis for a discussion on how control systems with high level selection criteria can exploit noise}.

From a theoretical point of view, this causation class is a dominant source of  top-down causation in financial markets. Deviations away from asset prices determined by the shared-risk factors are assumed to be sufficiently small to become negligible and correlations in the noise, due to activities lower in the hierarchy, are also regarded as negligible. This means that the direct impact on information variables, $Z_{k,t}$ or $\theta_{i,m,t}$ are either random or have short-lived effects and  that only aggregates effects are of concern for  risk variables $r_{p,t}$. This may  drive asset prices to cluster in groups via the shared risk-factors when capital is deployed based on those shared-risks.


\subsection{Top-down causation via feedback control of adaptive goals} \label{sec:trade}

This class of causality is characterised by adaptively selected goals, which respond to context and are used to guide the outcomes \cite{GFRE2008}. Consider, for example, market participants who incorporate information about demand and random fluctuations, to offset risks for more optimal trading. This can be carried out by implementing different trading strategies, contingent on prevailing conditions.

{\bf Example: Traders:} Participants in the market which may serve as an exemplar for this causation class include  market-makers, who facilitate transactions at all times,  and traders, who interact with the market on behalf of clients for fees.

The dynamic nature of goals, conditioned on past information, may cause groups of agents to herd based on common objectives \cite{CL2013}.
Here the behaviours of agents are based on the  information variables,
$Z_{k,t}$ and $\theta_{i,m,t}$, and shared-risks $r_{p,t}$, with goals updated based on past states of the system as well.
We note that this type of causation may include both bottom-up and top-down effects. Goals may be selected to guide outcomes based on properties of the system which reside higher in the hierarchy, for example, a trader may observe the spread between a pair of tradeable securities and make trading decisions in terms of price levels, while at the same time using her balance sheet to influence and move the prices themselves to trigger or action ad hoc outcomes. Noise variables $\eta_{s_i}$ may become significant if correlations in the noise become dominant.

\subsection{Top-down causation with adaptive selection of adaptive goals} \label{sec:reg}

{\bf Examples: Regulators and Rulers: } Regulators attempt to act on a financial market based on the intelligent and reasonable formulation of rules. For example, changing the market micro-structure at the lowest level in the hierarchy, can change the way that asset prices assimilate changes in information variables $Z_{k,t}$ or $\theta_{i,m,t}$.
Similarly, changes in accounting rules could change the meaning and behaviour of bottom-up information variables $\theta_{i,m,t}$ and changes in economic policy and policy implementation can change the meaning of top-down information variables $Z_{k,t}$ and influence shared risk factors $r_{p,t}$.

In hierarchical analysis, theories and plans may be embodied in a symbolic system to build effective and robust models to be used for detecting deeper dependencies and emergent phenomena\cite{Aubin2003,Minsky2008, Haldane2012, Helbing2013}. Mechanisms for the transmission of information and asymmetric information information have impacts on market quality\cite{Akerlof1970,Spence1973,GS1976,Stiglitz2000,OP2007,OY2011, WG2013b,PIS2014}.  Thus, Regulators can impact the activity and success of all the other actors, either directly  or indirectly through knock-on effects. Examples include the following:  Investor behaviour could change the goal selection of Traders; change in the latter could in turn impact variables coupled to  Traders activity in such a way that Profiteers are able to benefit from change in liquidity or use leverage as a mean to achieve profit targets and overcome noise.

 Idealistically, Regulators may aim  for  increasing productivity, managing inflation, reducing  unemployment and eliminating malfeasance.
However, the circumvention of rules, usually in the name of innovation or by claims of greater insight on optimality, is as much part of a complex system in which participants can respond to rules.  Tax arbitrages are examples of actions which manipulate reporting to reduce levies  paid to a profit-facilitating system. In regulatory arbitrage, rules may be followed technically, but nevertheless  use relevant new information which has not been accounted for in system rules. Such activities are consistent with goals of profiteering but are not necessarily in agreement with longer term optimality of reliable and fair markets.

Rulers, i.e. agencies which control populations more generally,  also impact markets and economies. Examples of top-down causation here include segregation of workers and differential assignment of economic rights to market participants, as in the evolution of local miners' rights in the late  1800's in South Africa and the national Native Land act of 1913 in South Africa, international agreements such as the Bretton Woods system, the Marshall plan of 1948,   the lifting of the gold standard in 1973 and  the regulation of capital allocations and capital flows between individual and aggregated participants.

 Ideas on target-based goal selection are already in circulation in the literature on applications of viability theory and stochastic control in economics\cite{A2004,OS2014}. Such approaches   provide alternatives to the Laplacian ideal of attaining perfect prediction by offering analysable  future expectations to regulators and rulers.

%

\section{Conclusion}

In order to address the need for more realistic economic models which include feedbacks, adaptive goal-seeking, emergence and the occurrence of crises, we have considered the applicability of a particular hierarchical modelling approach \cite{GFRE2008} by considering specific model interactions.

In Section IV, we provided a consistent set of coupled models, incorporating causation at differently levels in the hierarchy, driven by top-down sources, realised as {\em actors},  and bottom-up information variables.  The model incorporates the possibility of emergence, via random interactions, but is still consistent with no-arbitrage equity models at relevant scales, and can be calibrated for forecasting, based on historic data.
We find the  actor-based perspective of causation sources to be useful for identifying feedbacks and non-trivial features, which influence whether the system is in fact fit-for-purpose, in the sense of having low-coupling and high-cohesion. Here, insights can be derived from both the qualitative features of actors and actor interactions as well as the quantitative outputs of simulations.

Since economies can only be run once, unlike small experiments in controlled laboratory conditions, it becomes completely relevant to scrutinise as much information as possible when stipulating and enforcing (or actively not enforcing) the economic impact of laws. While, no human-constructed model is ever likely to include all significant information and dynamics, increased computing power, advances in complex systems modelling and better quantification of system information may yield  insight-producing simulations for better economic decision making.

The specific instantiation of a hierarchical model discussed in this paper is by no means a unique solution to the challenge of finding  relevant hierarchical models in financial economics. Different actors may impact different components of the system and more than one theory may be effective for a given level in the hierarchy under scrutiny. Thus, the approach is comprehensive and pluralistic, provided consistency constraints are maintained at the interfaces between models, for investigating multi-layer models in finance.

\begin{acknowledgments}
The authors thank George Ellis, Franco Bussetti, Antoinette and Robert Wilcox, Ron Cross, Dieter Hendricks and Virginie Konlack for discussions on complexity and financial markets. This work is based on the research supported in part by the
National Research Foundation of South Africa (Grant numbers
87830, 74223 and 70643). The conclusions herein are due to
the authors and the NRF accepts no liability in this regard
\end{acknowledgments}


\begin{thebibliography}{100}
\bibitem{foot1}This form of the price model ensures that the self correlation of a stock is one and independent of the cluster coupling. This can be seen by computing the self correlation $E[X_i^2]$ and using that clusters and stock unique process are unit variance zero mean processes:
\begin{eqnarray}
\mathrm{E}[(g_{s_i} \eta_{s_i} + \sqrt{1-g_{s_i}^2} \epsilon_i)^2] = g_{s_i}^2 + (1-g_{s_i}^2) = 1.
\end{eqnarray}
This is not a unique choice, another possible choice often used is:
\begin{eqnarray}
\mathrm{E}[(\frac{\sqrt{g_{s_i}}}{\sqrt{1+g_{s_i}}} \eta_{s_i} + \frac{1}{\sqrt{1+g_{s_i}}} \epsilon_i)^2] = \frac{1 + g_{s_i}}{1+g_{s_i}} = 1.
\end{eqnarray}
\end{thebibliography}
\end{document}